\documentclass{baltzer}
\begin{document}
\input{psfig}
\begin{frontmatter}  
%
\title{Positron-Neutrino Correlations in $^{32}$Ar and $^{33}$Ar Decays: 
Probes of Scalar Weak Currents and Nuclear Isospin Mixing} 
\author[A]{A. Garc\'{\i}a,}
\author[B]{E.G. Adelberger,}
\author[A]{C. Ortiz,}
\author[B]{H.E. Swanson,}
\author[B]{M. Beck,}
\author[C]{O. Tengblad,}
\author[C]{ M.J.G. Borge,}
\author[D,E]{I. Martel,}
\author[B]{H. Bichsel,}
\author[D]{and the ISOLDE Collaboration.}
\address[A]{University of Notre Dame, Notre Dame, IN 46556
\email{garcia.37@nd.edu}}   
\address[B]{University of Washington, Seattle, WA 98195
\email{eric@npl.washington.edu}}
\address[C]{Instituto de Estructura de la Materia,
CSIC, E-28006 Madrid, Spain}
\address[D]{EP Division, CERN, Geneva, Switzerland CH-1211}
\address[E]{Escuela Politecnica Superior, 21071 Huelva, Spain}
\runningauthor{}
\runningtitle{}
%
\begin{abstract}  
The positron-neutrino correlation in the $0^+ \rightarrow 0^+$
$\beta$ decay of $^{32}$Ar was measured at ISOLDE by analyzing the effect
of lepton recoil on the shape of the narrow proton group following
the superallowed decay.
Our result is consistent with the Standard Model prediction;
for vanishing Fierz interference we find
$a=0.9989 \pm 0.0052 \pm 0.0036$.
Our result leads to improved constraints on scalar weak interactions.
The positron-neutrino correlation in $^{33}$Ar decay was measured in the same 
experiment; for vanishing Fierz interference we find $a=0.944 \pm 0.002
\pm 0.003$.
The $^{32}$Ar and $^{33}$Ar correlations, in combination with precision 
measurements of the half-lives, superallowed branching ratios and beta 
endpoint energies, 
will determine the isospin impurities of the superallowed transitions.
These will provide useful tests of isospin-violation corrections
used in deducing $|V_{\rm ud}|$ which currently indicates 
non-unitarity of the KM matrix.
\end{abstract}
\end{frontmatter}
\section{The $(e,\nu)$ Correlation as a Probe of
Physics Beyond the Standard Model}
\label{sec: intro}

In the Standard Model, nuclear $\beta$ decay is mediated by the exchange 
of W bosons which have only vector and axial-vector couplings. However,
extensions of the standard model, such as super-symmetric theories
with more than one charged Higgs doublet, or leptoquarks, naturally
predict scalar or tensor weak couplings~\cite{he:95}.
A general effective Hamiltonian for allowed $\beta$ transitions
that respects Lorentz invariance is~\cite{ja:57}
\begin{eqnarray}
H = (\bar{\psi_p} \gamma_\mu \psi_n)
         (C_V  \bar{\psi_e} \gamma_\mu \psi_\nu +
          C_V' \bar{\psi_e} \gamma_\mu \gamma_5 \psi_\nu) \nonumber\\
    +(\bar{\psi_p} \gamma_\mu \gamma_5 \psi_n)
         (C_A  \bar{\psi_e} \gamma_\mu \gamma_5 \psi_\nu +
          C_A' \bar{\psi_e} \gamma_\mu  \psi_\nu)  \nonumber\\
    +(\bar{\psi_p} \psi_n)
         (C_S  \bar{\psi_e} \psi_\nu +
          C_S' \bar{\psi_e} \gamma_5 \psi_\nu) \nonumber\\
    +\frac{1}{2}(\bar{\psi_p} \sigma_{\lambda\mu} \psi_n)
         (C_T  \bar{\psi_e} \sigma_{\lambda\mu} \psi_\nu +
          C_T' \bar{\psi_e} \sigma_{\lambda\mu} \gamma_5 \psi_\nu) \nonumber\\
	    ~ + {\rm Hermitian~conj.}
\label{hamilt_eq}
\end{eqnarray}
where a term proportional to $(\bar{\psi_p} \gamma_5 \psi_n)$
has been neglected because nucleons are non-relativistic.
In the Standard Model, 
$C_V = C_V'$, $C_A = C_A'$,
and $C_S = C_S' =C_T = C_T'= 0$. 
Jackson {\em et al.}~\cite{ja:57} computed the
nuclear-$\beta$-decay rate from this Hamiltonian; for an unoriented 
initial state and summation over the lepton helicities
\begin{eqnarray}
dW = dW_0 (1 + a {{\bf p_e} \cdot {\bf p_\nu} \over E_e E_\nu} +
b {m_e \over E_e} )~,
\label{eq: rate}
\end{eqnarray}
where
\begin{eqnarray}
a \xi = |M_F|^2 (|C_V|^2+|C_V'|^2-|C_S|^2-|C_S'|^2) \nonumber\\
-\frac{1}{3}
      |M_{GT}|^2 (|C_A|^2+|C_A'|^2-|C_T|^2-|C_T'|^2)
\label{eq: a}\\
b \xi =\pm2 \gamma  {\rm Re}[|M_F|^2(C_V^*C_S+C_V'^*C_S') \nonumber\\
                         +|M_{GT}|^2(C_A^*C_T+C_A'^*C_T')]
\label{eq: b}\\
{\rm and} \nonumber \\
 \xi = |M_F|^2 (|C_V|^2+|C_V'|^2+|C_S|^2+|C_S'|^2) \nonumber\\
   +|M_{GT}|^2 (|C_A|^2+|C_A'|^2+|C_T|^2+|C_T'|^2)~,
\end{eqnarray}
where $M_F$ and $M_{GT}$ are rank-0 and rank-1 nuclear matrix elements.
We have simplified the expressions by neglecting the small Coulomb 
effects, and adopting the allowed approximation (ignoring curvature
in the lepton wave functions and the recoil-order corrections). 
In what follows we shall assume $C_V=C_V^\prime$ and express 
the scalar couplings in terms of:
\begin{eqnarray}
\tilde{C_S}= C_S/C_V, ~~{\rm and}~~\tilde{C_S^\prime}= C_S^\prime/C_V.
\end{eqnarray}

Precise measurements of the $e$-$\nu$
correlation coefficient, $a$, and of the Fierz-interference term, $b$,
can potentially yield information about physics beyond
the standard model.
Equations 2-5 are complicated and depend on nuclear
physics information through $M_F$ and $M_{GT}$.
However, for pure Fermi or pure GT transitions
the expressions simplify and the $M_F$ and $M_{GT}$
factors cancel. In pure Fermi transitions, a positron-neutrino 
correlation
coefficient $a < 1$ would immediately imply 
the presence of scalar currents; while in GT decays a value
$a > -1/3$ would imply tensor currents.

To measure the $e$-$\nu$ correlation one must
determine the daughter's velocity, because it is out of the
question to detect
the low-energy neutrino. But, in general, measuring the 
daughter's recoil velocity is not a trivial task: the
velocities are very small and the interaction of the daughter with
the surrounding medium can jeopardize the measurement.
Fortunately nature has provided us with cases where
the daughter states are unbound to proton emission, so that the
daughter's velocity can be determined via the `Doppler'
broadening of the beta-delayed proton groups.
Protons are preferred over other $\beta$-delayed radiations because they
are emitted before the daughter nucleus has slowed appreciably and
can be detected with high resolution and good efficiency.
Neutrons, on the other hand, are hard to detect with high
resolution and high efficiency simultaneously. Gamma rays
are generally emitted too slowly to probe the daughter velocity
before it has been reduced by interaction with the
surrounding medium.
This paper discusses superallowed transitions
from proton-rich nuclei with proton-unbound daughter states.
These transitions are strong, which enhances the signal-to-noise 
ratio, and the widths of daughter states are ideally 
suited for probing the $e$-$\nu$ correlation:
\begin{itemize}
\item the time scale for proton decay is short enough so that
the decays occur before the daughter looses any appreciable
velocity due to interactions with the medium (in 
the superallowed decay of $^{32}$Ar, the daughter travels 
$\leq 2 \times 10^{-2}$ \AA~before the proton is emitted).
\item because proton decays in these cases violate isospin symmetry
the time scale for proton decay is long enough so the widths are
much narrower than the lepton-recoil broadening (in 
the superallowed decay of $^{32}$Ar, the daughter state has a
width of $\approx 20$ eV while the lepton-recoil broadening 
has a full-width at half-maximum of $\approx 25$ keV).
\end{itemize}

\section{Limits on Scalar Weak Interactions
from $^{32}$Ar $\beta^+$ Decay}
\label{sec: ar32}

We performed an experiment at ISOLDE that measured with 
high precision the energies of protons following the 
$0^+ \rightarrow 0^+$ $\beta^+$ decay of $^{32}$Ar.
Except for this experiment and a previous 
ISOLDE study of $^{32}$Ar by Schardt and Riisager~\cite{sc:93},
there are no other precise determinations of $e$-$\nu$
correlations in pure Fermi transitions, which explains why,
prior to our new result, 
limits on scalar couplings were rather 
poor in comparison to those on tensor currents~\cite{bo:84}.
We here describe the salient features of our experiment
and the extracted limits on scalar currents.
More details on the $^{32}$Ar experiment can be found in 
Ref.~\cite{ad:99}. 

Critical challenges for this kind of experiment are: 
\begin{enumerate}
\item obtaining an intense and pure source of radioactivity,
\item eliminating proton-$\beta^+$ summing, 
which distorts the shape of the proton peak,
\item and optimizing the energy resolution of the proton counter.
\end{enumerate}
ISOLDE solved problem 1 by producing very pure beams of both
$^{32}$Ar and $^{33}$Ar from a CaO target and a plasma ion source, 
providing an average of $\approx 94$ $^{32}$Ar's/s 
and $\approx 3900$ $^{33}$Ar's/s on our catcher 
foil over the 9-day-long run.
We solved problem 2 by immersing our detection system in a $3.5$ T
magnetic field.
The $^{32}$Ar beam from ISOLDE was stopped in a $\approx$23 $\mu$g/cm$^2$
C foil at 45 degrees to the beam. Our proton detectors were located at
$\pm 90$ degrees with respect to the beam and at about 1.6 cm from the beam
spot. In the 3.5 T field superallowed $\beta$'s had a 
maximum radius of $\approx 0.55$ cm,
while protons had $\approx 7.14$ cm.
We solved problem 3 by using cooled PIN-diode proton detectors 
($\approx 0.9 \times 0.9$ cm) and temperature-controlled electronics 
to obtain a proton energy resolution of $\approx 4.5$ keV
($\approx 3.0$ keV electronic noise). 
This greatly enhanced our sensitivity to the $e$-$\nu$ correlation.

\begin{figure}[ht]
\hspace{-1.2cm}
\hbox{\psfig{figure=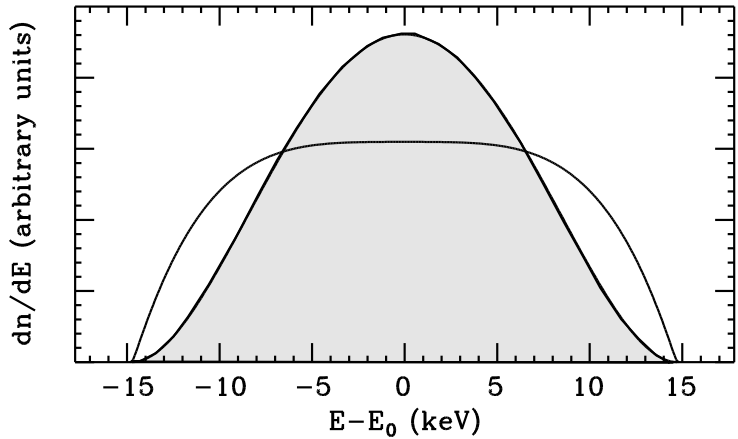,height=22cm}}
\vspace{-14cm}
\caption{Intrinsic shapes of the
delayed proton group from $^{32}$Ar $0^+ \rightarrow 0^+$ decay for
$a=+1,~b=0$ (unshaded curve) and $a=-1,~b=0$ (light-shaded
curve).} 
\label{fig: MC}
\end{figure}
\begin{figure}[ht]
\hspace{-1.2cm}
\hbox{\vspace{-10cm}
\psfig{figure=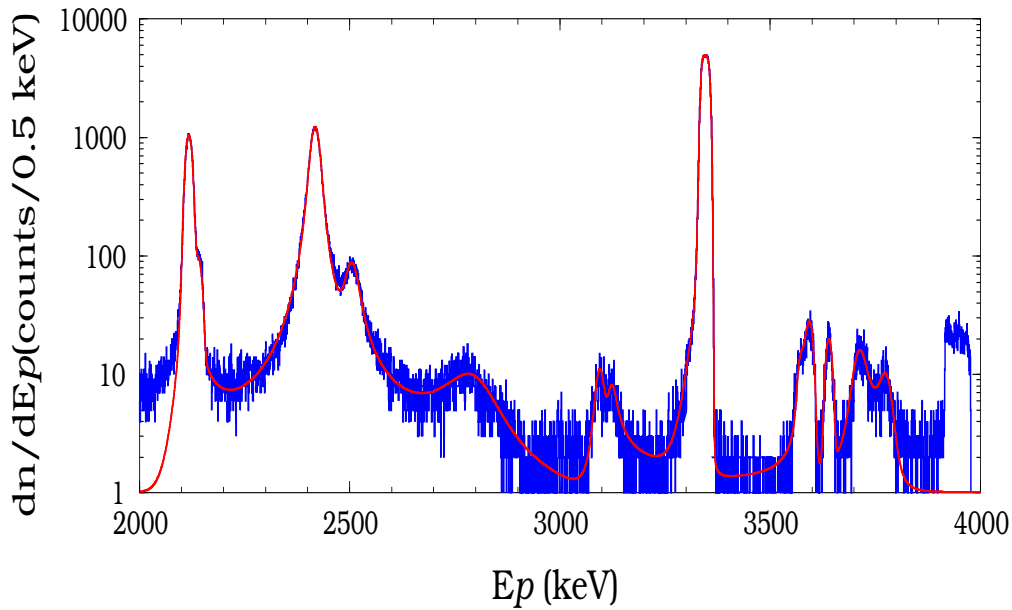,height=28cm,width=16cm}}
\vspace{-16cm}
\caption{R-matrix fit to the $^{32}$Ar delayed proton spectrum.
This spectrum, contains roughly 1/10 of
our data.}
\label{fig: fit_32ar}
\end{figure}
\begin{figure}[ht]
\hspace{-0.5cm}
\hbox{\psfig{figure=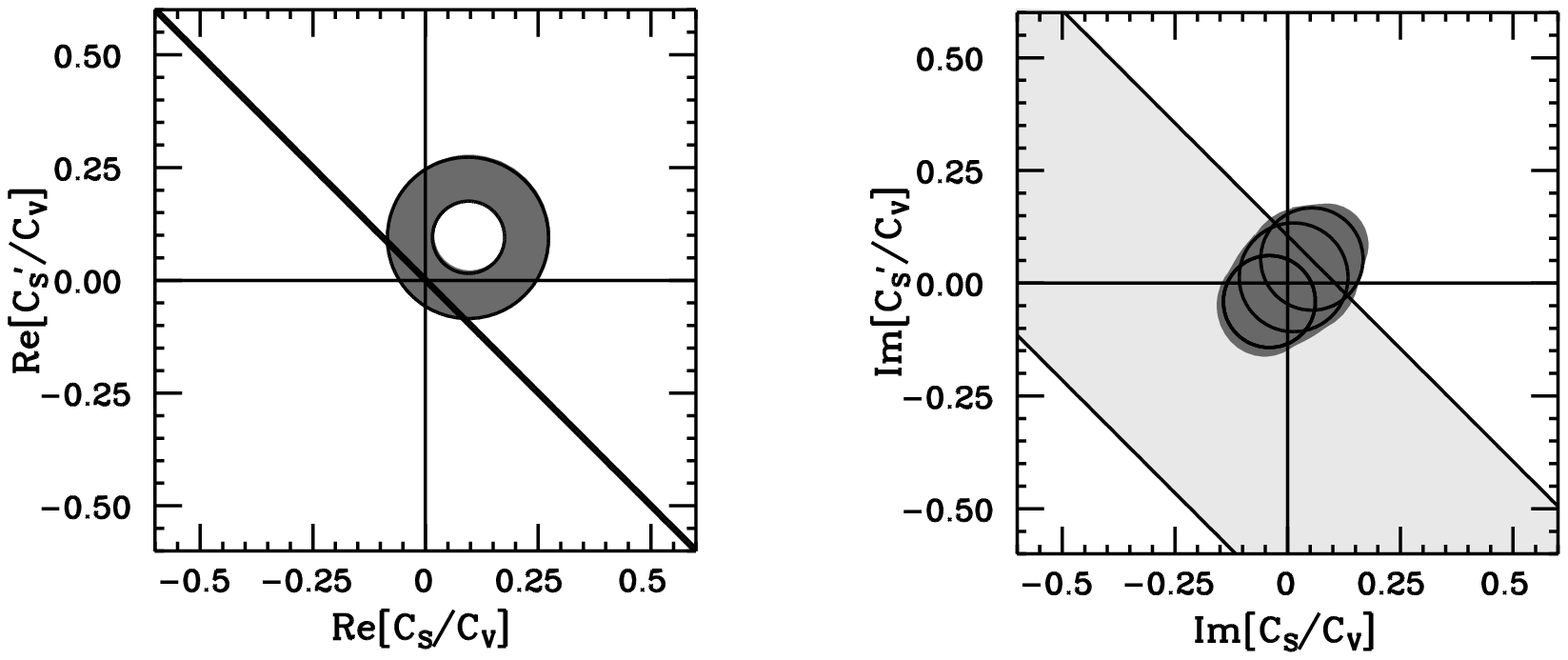,width=16cm}}
\vspace{-14cm}
\caption{95\% conf. limits on $\tilde{C}_S$ and $\tilde{C}_S^{\prime}$,
including statistical and systematic errors.
Left panel: time-reversal-even couplings. The annulus is from this work, 
the diagonal band is the Fierz interference result of
Ref.~\protect\cite{or:89}. Right panel: time-reversal-odd
couplings. The circles are from this work and correspond to
phases of $\tilde{C}_S$ and $\tilde{C}_S^{\prime}$ of
$\pm 90^{\circ}$, $+45^{\circ}$ and $-45^{\circ}$. The
shaded oval is the constraint with no assumptions about this phase.
The diagonal band is from the $R$-coefficient
in $^{19}$Ne decay \protect ~\cite{sc:83}.
}
\label{fig: cs_lim}
\end{figure}

Fig.~\ref{fig: MC} shows Monte-Carlo predictions for the
shape of the proton peak assuming the proton
detector had infinitely good energy resolution. 
If one assumes $b=0$ in Eq.~\ref{eq: rate}
one can extract $a$ by
producing linear combinations of these two shapes,
folding them with the detector response function 
(assumed to be two exponentials with adjustable tails and areas,
convoluted with a Gaussian of adjustable width) and determining
the values of the parameters that minimize $\chi ^2$.
Fig.~\ref{fig: fit_32ar} shows our best fit to a subset of our data 
containing
approximately 1/10 
of the statistics. The very good energy
resolution obtained in this study also allowed us to extract 
valuable spectroscopic information on $^{32}$Cl. 
The fit was performed using an R-matrix parametrization of the 
resonances. Reduced total widths, $\Gamma_p$, and 
ratios $\Gamma_{p1}/\Gamma_{p0}$ can be extracted for many 
of the resonances.
For the general case when $b \ne 0$ one has to fold an additional
distribution (taking into account the $m/E$ term in 
Eq.~\ref{eq: rate}). We produced a grid in the 
$\tilde{C_S},\tilde{C_S^\prime}$
space and, for each point, minimized $\chi^2$ with respect
to the response function parameters. The resulting confidence 
regions are shown in  Fig.~\ref{fig: cs_lim}. 
We found that our $\tilde{C}_S$, $\tilde{C}_S^{\prime}$ 
constraints are well reproduced by the single parameter
\begin{eqnarray}
\tilde{a} &\equiv& a/(1 + 0.1913 b)~ \nonumber \\
\label{eq: result}
\end{eqnarray}
where $a$ and $b$ are given in Eqs.~\ref{eq: a}, \ref{eq: b}.
In other words, replacing $m/E$ by an appropriate average
reproduces the regions of interest and allows us
to quote our results in terms of $\tilde{a}$.
Our experiment yields the constraint
\begin{eqnarray}
\tilde{a} = 0.9989 \pm 0.0052({\rm stat.}) \pm 0.0036({\rm syst.})~
{\rm 68\%~c.l.}
\end{eqnarray} 
The systematic error was evaluated by redoing the whole analysis
under different conditions. We found 
$\partial \tilde{a}/\partial \Delta = -1.2\times 10^{-3}$~keV$^{-1}$ 
where $\Delta$ is the $\beta$-decay endpoint; and
$\partial \tilde{a}/\partial Q_p=-0.9 \times 10^{-3}$~keV$^{-1}$,
where $Q_p$ is the energy of the emitted proton. 
We measured $Q_p$ with
an uncertainty, $\delta Q_p= \pm 1.2$~keV, by alternating between
$\sim 2$ h $^{32}$Ar runs with 10-15 min $^{33}$Ar runs that
gave us a continuous calibration of the energy scale. 
The mass of $^{32}$Ar has been determined
only to within 50 keV~\cite{au:95}, which would impose a systematic
error of $\approx 6$\% on our measurement. Fortunately, 
as shown in Table~\ref{tab: IMME32}, the masses of all other 
members of the $T=2$ isospin multiplet are known with high precision. 
\begin{table}
\begin{flushleft}
\caption{Comparison of the measured mass excesses of the lowest $T=2$
quintet in $A=32$ to predictions of the Isospin-Multiplet Mass
Equation [$P(\chi^2, \nu)=0.73$].}
\label{tab: IMME32}
\begin{tabular}{lrrr}
\hline
\hline
isobar & $T_3$ & $M_{\rm exp}$~(keV)$^a$ & $M_{{\rm IMME}}$
(keV) \\
\hline
$^{32}$Si        & $+2$ & $-24080.9 \pm 2.2$ & $-24081.9 \pm 1.4$ \\
$^{32}$P         & $+1$ & $-19232.88 \pm 0.20$$^b$ & $-19232.9 \pm 0.2$ \\
$^{32}$S         & $0$ & $-13970.98 \pm 0.41$$^c$ & $-13971.1 \pm 0.4$ \\
$^{32}$Cl        & $-1$ & $-8296.9\pm 1.2$$^d$ & $-8296.6 \pm 1.1$ \\
$^{32}$Ar        & $-2$ & $-2180 \pm 50$ & $-2209.3 \pm 3.2 $ \\
\hline
\end{tabular}
\\
$^a$unless noted otherwise, ground state masses are from 
Ref.~\protect\cite{au:95}.\\
$^b$$E_x=5072.44 \pm 0.06$ keV from Ref.~\protect\cite{en:90}.\\
$^c$$E_x=12045.0 \pm 0.4$ keV from Ref.~\protect\cite{an:85,wa:98}.\\
$^d$from delayed proton energy measured here and masses of
Ref.~\protect\cite{au:95}.\\
\end{flushleft}
\end{table}
We use the Isospin-Multiplet Mass Equation~\cite{an:85},
$M(T_3)=a +b T_3 +c T_3^2$, to obtain  
\begin{equation}
\Delta = 6087.3 \pm 2.2~{\rm keV}.
\end{equation}
\begin{figure}[ht]
\hspace{-0.5cm}
\hbox{\vspace{-6cm}
\psfig{figure=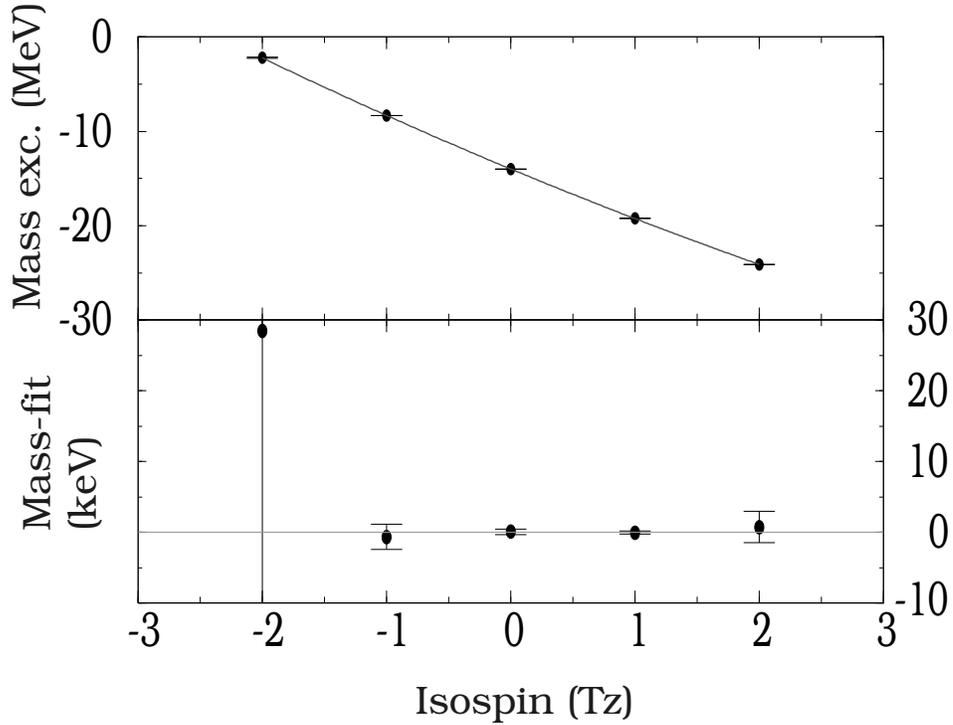,width=14cm}}
\vspace{-5cm}
\caption{Plot of residuals of IMME fit to the 
$T=2$ quintuplet in $A=32$. Top panel: mass excesses in MeV. 
The points show the measured values and the continuous line
shows the IMME fit.
Bottom panel: fit residuals in keV.}
\label{fig: IMME32}
\end{figure}
As shown in Fig.~\ref{fig: IMME32} we obtain an excellent fit 
to the IMME with $P(\chi^2,\nu)=0.71$. On the other hand, modifying the
IMME by adding a $dT_3^3$ term we obtain
$\Delta = 6086.7 \pm 4.9$ keV and $d =0.25\pm0.47$ keV, 
but with a lower probability, $P(\chi^2,\nu)=0.52$, indicating that there
is no empirical basis for adding a term to the IMME. 
We expect the IMME to work very well for the $A=32$ multiplet where
the states are relatively well-bound and the Coulomb barriers 
relatively high. 
Small non-zero $d$ terms 
(never more significant than $3 \sigma$) have been observed in 
light nuclei where the $T_3=T$ members are much closer 
to being unbound (or even unbound) and the Coulomb barriers much lower. 
We here assume $\delta \Delta=\pm 2.2$~keV
which combined with the uncertainty in $Q_p$ yields a
{\em kinematic} systematic error $\delta \tilde{a} = \pm 0.0032$.
Note that even when $d$ is allowed to vary freely 
the uncertainty is only about twice this value.
We also checked the dependence of $\tilde{a}$ on the 
{\em fitting regions} of the proton spectra; a 28\% variation in the
width of the region changed $\tilde{a}$ by less than $\pm 0.00055$.
We examined the dependence of our results
on the form of the detector response function by re-analysing
the data with a single-tail response function;
by re-analyzing the data assuming that
a weak Gamow-Teller peak lay under the tail of the $^{32}$Ar superallowed
peak; and by simultaneously fitting the
$^{33}$Ar and $^{32}$Ar superallowed peaks using a common response
function. From these tests we inferred a
{\em line-shape} systematic error of $\delta\tilde{a} = \pm 0.0016$.

\begin{figure}[ht]
\hspace{-0.5cm}
\vspace{2cm}
\hbox{\psfig{figure=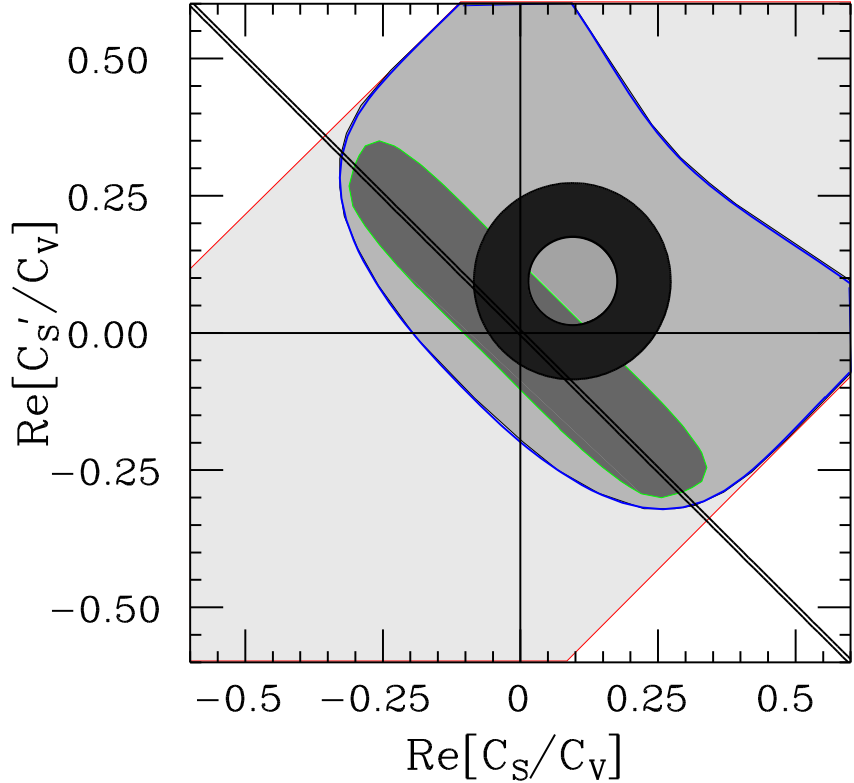,width=16cm}}
\vspace{-10cm}
\caption{Comparison of our constraints on scalar couplings with 
previous work. The light shaded area represents constraints from
neutron $\beta$ decay alone: {\em i.e.} from measurements of $a$,
 $A$, $B$, and $t_{1/2}$~\cite{pdg:98}. 
A slightly darker area shows how these 
constraints improve when combined with measurements of
the polarization of $\beta$'s from $^{14}$O and 
$^{10}$C~\cite{ca:91}.
The darker shaded area shows the result of adding to the 
previous the constraints on Fierz terms from $^{22}$Na~\cite{we:68}
and the measurement of $a$ in $^6$He~\cite{jo:63}.
The darkest shaded area shows constraints 
from our results. The narrow area looking like a line at 
$-45 ^\circ$ is from constraints on Fierz terms from 
$0^+ \rightarrow 0^+$ transitions.}
\label{fig: comparison} 
\end{figure}
Figure~\ref{fig: comparison} compares our results to
previous constraints on scalar couplings. 
For scalar interactions with
$\tilde{C}_S=-\tilde{C}_S^{\prime}$ so that $b=0$,
our data yield the $1\sigma$ constraint
$|\tilde{C}_S|^2 \leq 3.6 \times 10^{-3}$. The corresponding lower
limit on the mass of scalar particles with gauge coupling strength
is $M_S = |\tilde{C}_S|^{-1/2} M_W \geq 4.1 M_W$.
We note that data from neutron $\beta$ decay by itself
does not place stringent constraints on scalar couplings
because the measurement on the correlation coefficient
is not very accurate ($\delta a/a \approx 5$\%)
and neutron decay is sensitive to both scalar and 
tensor couplings. 
Moreover, because of the
larger value of $<m/E>$ ({\em i.e.} the lower endpoint energy),
the circle generated by the equivalent of Eq.~\ref{eq: result}
has a large radius which weakens the neutron constraints. Even when 
supplemented by other data on the GT Fierz interference {\em etc.},
the neutron constraints are not as tight as those from 
the present work and limits on Fierz interference in
$0^+ \rightarrow 0^+$ transitions.

\section{Isospin mixing Corrections to $0^+ \rightarrow 0^+$
${\cal F}t$ values and $|V_{ud}|$}

Taken at face value, the $V_{ud}$ matrix element extracted from the
${\cal F}t$ values of nine $0^+ \rightarrow 0^+$
$\beta$-decay transitions
implies non-unitarity of the Kobayashi-Maskawa 
matrix~\cite{ha:98}. 
Because of the importance and unexpected nature of this conclusion,
it is worth reexamining whether any systematic effect could affect the
${\cal F}t$ values.
One possibility concerns the 
corrections for isospin-symmetry violation in the 
parent and daughter nuclear wave functions.
These corrections are usually separated
into `configuration mixing' and `nucleon overlap' parts;
the latter being dominant ($\approx $ four times the 
former).
Although several authors ~\cite{to:77,or:95,sa:96} 
have performed
independent calculations that agree reasonably well
it would be valuable to check these calculations
on additional transitions in neighboring nuclei.

The cases we present below provide a good opportunity to check the 
calculated corrections; the `nucleon overlap' corrections are
enhanced over those in the nine standard cases because the nuclei lie farther
from the valley of stability.

\subsection{Isospin Mixing in the Fermi decay of $^{32}$Ar}
The isospin-mixing correction in $^{32}$Ar was calculated by
B.A. Brown~\cite{br:98} using the SKX Skyrme interaction~\cite{br:98c}.
This calculation yields $2.0 \pm 0.4$\%  where the uncertainty is based on
previous comparisons of similar calculations to measurements.
The large size of the correction
is due to the looser binding of the $d \frac{3}{2}$ and
$s \frac{1}{2}$ proton states compared to the neutron states. 
For comparison, the average correction for the nine 
standard cases is $\approx 0.41$\%, while the correction for 
the neighboring decay of $^{34}$Cl is $\approx 0.61$\%.

In $^{32}$Ar decay there is no mixing with GT transitions so that
one can check the isospin-mixing correction simply by  
determining the half-life of the parent, and the branching ratio and 
endpoint of the superallowed transition. 
One can thus extract the ${\cal F} t$ value and compare with 
the prediction.
Furthermore,
if some Fermi strength is diverted into narrow $J^{\pi}=0^+;~T=1$ 
levels in the 
$^{32}$Cl daughter, it may be possible to identify these
transitions because they will have $a = +1$ instead of $a=-1/3$.

We expect to determine
the $^{32}$Ar half life to $\approx 0.2$\% from our ISOLDE experiment 
(our preliminary value is $t_{1/2}=100.74 \pm 0.18$ ms) and we hope to 
determine the branching ratio with high precision at an upcoming experiment 
at MSU~\cite{ko:99}.

The extraction of the $^{32}$Ar mass from the IMME presented
in section~\ref{sec: ar32} yields the endpoint to $\approx 2.2$ keV,
which allows to calculate the phase space factor to $\approx 0.3$\%.
The $^{32}$Ar mass could also be measured at ISOLTRAP~\cite{bo:98}
but due to the short $^{32}$Ar half-life it is not clear that 
one could get enough intensity to determine the endpoint to within
a few keV. 
All of the above indicates that one could extract the ${\cal F} t$
value to $\approx 0.41$\%.
\subsection{Isospin Mixing in the Fermi decay of $^{33}$Ar}
The theoretical isospin-mixing calculations can also be tested by
studying the superallowed decays of the $A=4n+1$ nuclei. These
are mixed Fermi/GT transitions so that one needs to 
determine the $B({\rm F})/B({\rm GT})$ ratios as well as the
half-lives, branching ratios and energy releases of the superallowed
transitions. The $B({\rm F})/B({\rm GT})$ ratio
can be obtained from the positron-neutrino correlation.
As pointed out in Section~\ref{sec: intro} the $e$-$\nu$
correlation from mixed Fermi-GT transitions is not as useful
for extracting information on scalar or tensor currents because
one needs to know the relative amounts of each component.
However, one can take the existing limits on scalar and tensor
currents from other experiments
and use the  $e$-$\nu$ correlation to extract the 
$B$(GT)/$B$(F) ratio. For simplicity, we here assume 
the scalar and tensor couplings to be zero.

We will soon gather the necessary information for the case of $^{33}$Ar.
We will obtain the half-life of $^{33}$Ar and the
absolute branching ratio of the superallowed
transition from our experiment at MSU~\cite{ko:99}.
The endpoint can be extracted from the Isospin-Multiplet Mass Equation.
\begin{table}
\begin{flushleft}
\caption{Comparison of the measured mass excesses of the lowest $T=3/2$
quintet in $A=33$ to predictions of the Isospin-Multiplet Mass
Equation [$P(\chi^2, \nu)=0.51$].}
\label{tab: IMME33}
\begin{tabular}{lrrr}
\hline
\hline
isobar & $T_3$ & $M_{\rm exp}$~(keV)$^a$ & $M_{{\rm IMME}}$
(keV) \\
\hline
$^{33}$P         & $+3/2$ & $-26337.7  \pm 1.1 $ & $-26337.7\pm 1.1$ \\
$^{33}$S         & $+1/2$ & $-21106.14 \pm 0.41$$^b$ & $-21106.15\pm 0.41$ \\
$^{33}$Cl        & $-1/2$ & $-15460.2  \pm 1.0 $$^c$ & $-15460.10 \pm 1.0$ \\
$^{33}$Ar        & $-3/2$ & $-9380  \pm 30$ & $-9399.54 \pm 3.4 $ \\
\hline
\end{tabular}
\\
$^a$unless noted otherwise, ground state masses are from 
Ref.~\protect\cite{au:95}.\\
$^b$excitation energy from Ref.~\protect\cite{en:90} and
masses of Ref.~\protect\cite{au:95}.\\
$^c$from $^{32}$S($p,p$) resonance energy\cite{en:90} 
and masses of Ref.~\protect\cite{au:95}.\\
\end{flushleft}
\end{table}
Table~\ref{tab: IMME33} shows the mass excesses for the $T=3/2$ 
quartet and the corresponding values of the IMME fit. Here we get:
\begin{equation}
\Delta(^{33}{\rm Ar})= 6060.6 \pm 2.6 ~{\rm keV.}
\end{equation}
In addition there are plans to determine the $^{33}$Ar mass
to $\approx 3$ keV using ISOLTRAP\cite{bo:98}
(the mass of the daughter level is known to $\approx 1$ keV).
\begin{figure}[ht]
\hspace{-1.2cm}
\hbox{\vspace{-4cm}
\psfig{figure=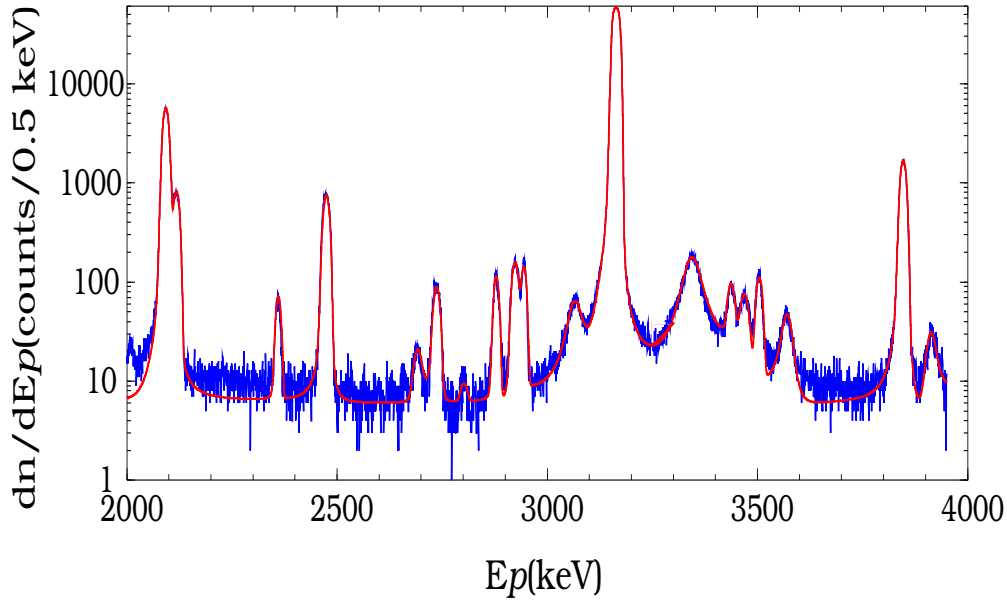,height=28cm,width=16cm}}
\vspace{-15cm}
\caption{R-matrix fit to the $^{33}$Ar delayed proton spectrum.
This spectrum, contains roughly 1/10 of our data.}
\label{fig: fit_33ar}
\end{figure}

We obtain $a$ from the $^{33}$Ar delayed proton spectrum, a typical example of
which is shown in Fig~\ref{fig: fit_33ar}. Our fit to the ISOLDE data 
yields
\begin{equation}
a(^{33}{\rm Ar})= 0.944 \pm 0.002({\rm stat.}) \pm 0.003({\rm syst.}), 
\label{eq: a_33_ours}
\end{equation}
which implies 
$B({\rm GT})/B({\rm F})=0.044 \pm 0.002$~\cite{note1}.
Our result is in reasonably good agreement with the
the shell-model calculation of ~\cite{br:85} which predicts
$B({\rm GT})/B({\rm F}) = 0.055$,
but in strong disagreement with the previous determination
of Schardt and Riisager who obtained~\cite{sc:93}
\begin{equation}
a(^{33}{\rm Ar}) > 1.02 \pm 0.04 \\
\label{eq: a_33_sr}
\end{equation}
($2 \sigma$ error bars), 
which can be translated into an upper limit 
$B({\rm GT})/B({\rm F}) < 0.015 $
which is $\approx 4$ of their $\sigma$'s from our value.

We evaluated the systematic errors in $a$ following the same
procedure we used for $^{32}$Ar. 
Using the endpoint deduced from the IMME and
the masses of Table~\ref{tab: IMME33},
combined with the known $Q_p = 2276.5 \pm 1.0$ keV,
and the derivatives 
$\partial \tilde{a}/\partial \Delta = -9.1 \times 10^{-4}$~keV$^{-1}$; 
and 
$\partial \tilde{a}/\partial Q_p=-8.5 \times 10^{-4}$~keV$^{-1}$,
extracted by redoing the analysis with different values of $\Delta$ and
$Q_p$, yields $\delta \tilde{a} = 0.0023$. 
Varying the width of the fitting region by 28\% 
yields variations of $\delta \tilde{a} \approx 0.0004$. 
Simultaneously fitting the $^{32}$Ar and $^{33}$Ar data
with a common detector response function
yields $\delta \tilde{a} \approx 0.002$.
These three uncertainties were combined in quadratures
to give the total systematic error
shown in Eq.~\ref{eq: a_33_ours}.

The discrepancy between our result in Eq.~\ref{eq: a_33_ours}
and that of Schardt and Riisager in Eq.~\ref{eq: a_33_sr} is due primarily
to differences in the analysis rather than disagreement of the data itself.
Dieter Schardt kindly made the raw data of Ref.~\cite{sc:93} available to us
and our analysis of their data gave a result essentially consistent with 
Eq.~\ref{eq: a_33_ours}.

We now can show the potential value of measuring the 
${\cal F} t$ value for the
superallowed transition by imagining that the total
strength $B({\rm F})+B({\rm GT})$ has been determined to 
$\approx 0.3$\%. 
Combining this with 
our positron-neutrino correlation measurement would
yield $B({\rm F})$ to $\approx 0.5$\%. 
The predicted~\cite{br:98} isospin-mixing correction for $^{33}$Ar
is $\approx 1.2$\%. For comparison, the correction in $^{34}$Cl
is $\approx 0.6$\%. The fact that the correction is enhanced, 
as in $^{32}$Ar, makes it easier to 
measure. So these measurements could check whether 
the calculated corrections are accurate to within 
$\approx$50\%. 
It should be noted that larger discrepancies 
(corrections should be $\approx 0.7$\% as 
opposed to $\approx 0.4$\%~\cite{ha:98a})
would be needed to explain away the 
apparent non-unitarity of the Kobayashi-Maskawa matrix. 
\\
\\
\noindent
{\bf Acknowledgments}\\
We thank D. Forkel-Wirth for help setting up our apparatus at CERN.
This work was supported in part by the 
USA National Science Foundation and the Warren Foundation
(at the University of Notre Dame) and by
the Department of Energy (at the University of Washington).

%

\newpage
\begin{thebibliography}{00} 
%
\bibitem{he:95} P. Herczeg, in {\em Precision Tests of the Standard
Electroweak Model}, World Scientific, P. Langacker ed., 768 (1995).
%
\bibitem{ja:57}J.D. Jackson, S.B. Treiman, and H.W. Wyld Jr.,
Nucl. Phys.  {\bf 4}, 206 (1957).
%
\bibitem{sc:93}D. Schardt and K. Riisager, 
Z. Physik A {\bf 345}, 265 (1993).
%
\bibitem{bo:84}A.I. Boothroyd, J. Markey, and P. Vogel,
Phys. Rev. C, {\bf 29} 603 (1984).
%
\bibitem{ad:99}E.G. Adelberger, C. Ortiz, A.~Garc\'{\i}a, 
H.E. Swanson, M. Beck, O. Tengblad, M.J.G. Borge, 
I. Martel-Bravo, H. Bichsel, and the ISOLDE collaboration,
submitted to Phys. Rev. Lett. 
Los Alamos Print Archive: nucl-ex9903002.
%
\bibitem{or:89}W.E. Ormand, B.A. Brown and B.R. Holstein, 
Phys. Rev. C {\bf 40}, 2914 (1989).
%
\bibitem{sc:83}M.B.~Schneider {\em et al.},
Phys. Rev. Lett. {\bf 51}, 1239 (1983).
%
\bibitem{au:95} G. Audi and A.H. Wapstra, 
Nucl. Phys. {\bf A 595}, 409, (1995).
%
\bibitem{en:90}P.M. Endt, 
Nucl. Phys. {\bf A521}, 1 (1990).
%
\bibitem{an:85} M.S. Antony {\em et al.}, 
At. Data Nucl. Data Tables {\bf 33}, 447 (1985).
%
\bibitem{wa:98}G. Walter, 
private communication (1998).
%
\bibitem{pdg:98} Particle Data Group, 
Eur. Phys. J. {\bf C 3}, 622 (1998).
%
\bibitem{ca:91}A.S. Carnoy {\em et al.}, 
Phys. Rev. C {\bf 43}, 2825 (1991).
%
\bibitem{we:68} H. Wenninger, J. Stiewe, and H. Leutz,
Nucl. Phys. {\bf A 109}, 561 (1968).
%
\bibitem{jo:63}C.H. Johnson, F. Pleasanton and T.A. Carlson, 
Phys. Rev. {\bf 132}, 1149 (1963).
%
\bibitem{note1} We define $B({\rm GT})$ as
\begin{eqnarray}
B({\rm GT})=g_A^2 {<J_f||{\cal O} ||J_i>^2 \over 2J_i+1} 
\protect \nonumber.
\end{eqnarray}
%
\bibitem{ha:98}E. Hagberg, J.C. Hardy, V.T. Koslowsky,
G. Savard, and I.S. Towner,
Los Alamos Print Archive: nucl-ex9609002.
%
\bibitem{to:77} I.S. Towner, J.C. Hardy, and M. Harvey,
Nucl. Phys. {\bf A284}, 269 (1977).
%
\bibitem{or:95}W.E. Ormand and B.A. Brown,
Phys. Rev. C {\bf 52} 2455 (1995).
%
\bibitem{sa:96}H. Sagawa, N. van Giai and T. Suzuki, 
Phys. Rev. C {\bf 53} 2163 (1996).
%
\bibitem{br:98}B.A. Brown, 
private communication.
%
\bibitem{br:98c}B.A. Brown, 
Phys. Rev. C {\bf 58} 220 (1998).
%
\bibitem{ko:99}A. Komives, A. Garc\'{\i}a, C. Ortiz, M. Bhattacharya,
E.G. Adelberger, H.E. Swanson, T. Glasmacher, P. Mantica, and B. A. Brown,
Experiment approved to run at MSU.
%
\bibitem{bo:98}G. Bollen, 
private communication.
%
\bibitem{br:85}B.A. Brown and B.H. Wildenthal, 
Atomic Data and Nucl. Data Tables {\bf 33}, 347 (1985).
%
\bibitem{ha:98a}J.C. Hardy and I.S. Towner,
ENAM 98, AIP Conference Proceedings 455,
page 733 (1998);
edited by B.M. Sherrill, D.J. Morrissey, and C.N. Davids.
%
\end{thebibliography}
\end{document}